\documentclass[12pt,a4paper]{article}

\author{E. Minguzzi \\ \small{{\em Dipartimento di Fisica dell'Universit\`a di Milano-Bicocca and}} \\ \small{{\em INFN, Sezione di Milano, via Celoria 16, 20133 Milano, Italy}} \\ \small{{\em  e-mail: minguzzi@mib.infn.it}}}
\title{ \bf ON THE CONVENTIONALITY  OF SIMULTANEITY}
\date{}
\usepackage[dvips]{graphicx}
\usepackage{amsthm}
\usepackage{amstext}
\usepackage{amsmath}
\usepackage{amsfonts}
\newcommand{\p}{\partial}

\newcommand{\dd}{{\rm d}}

\begin{document}
\maketitle \vspace{1cm}

 \noindent
Starting from the  experimental fact that light propagates over a
closed path  at  speed $c$ ($L/c$ law), we show to what extent the
isotropy of the speed of light  can be considered a matter of
convention. We prove the consistence of anisotropic and
inhomogeneous conventions,  limiting the allowed possibilities.
All conventions lead to the same physical theory even if its
formulation can change in form. The mathematics involved is that
of gauge theories and the choice of a simultaneity convention is
interpreted as a choice of the gauge. Moreover, we prove that a
Euclidean space where the $L/c$ law holds, gives rise to a
spacetime with Minkowskian causal structure, and we exploit the
consequences for the causal approach to the conventionality of
simultaneity.

\section{INTRODUCTION}

Since its birth \cite{HP,AE}, there has been a long debate to
establish to what extent special relativity, and the hypothesis of
constancy of the speed of light, could be considered conventional
\cite{RE,GR}. It was soon realized by Einstein \cite{AE} that
experimentally one can establish the speed of light only by
measuring the time of flight of a light beam over a closed path.
Indeed, in order to measure the {\em one-way} speed of light, the
time of departure from a starting point $O_{1}$ and the time of
arrival to an ending point $O_{2}$ are needed. A convention to
synchronize two distant clocks must be given, but the Einstein
convention cannot be used, since it is based on the isotropic
assumption which is the fact we wish to prove.

\begin{figure}[!ht]
\centering
\includegraphics{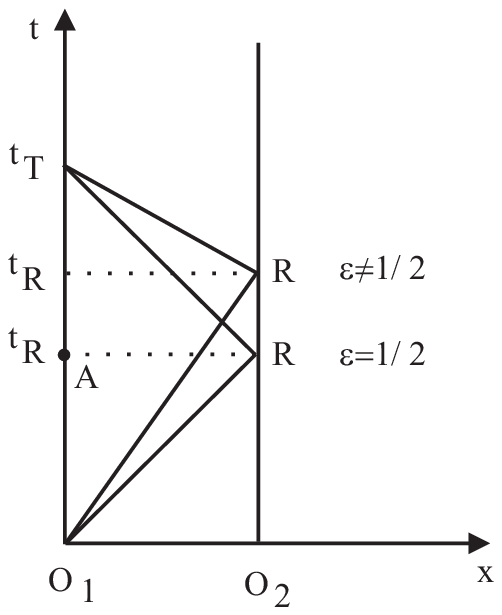}
\caption{Different values of $\epsilon$ give different definitions of simultaneity.}
\label{figura1}
\end{figure}

The situation is often illustrated, in the one dimensional case,
in the following way \cite{RE}(see figure \ref{figura1}).
Let us consider a light beam: it leaves the observer $O_{1}$,
reaches $O_{2}$ and, being reflected, it comes back. $O_{1}$, with
his clock, measures the total time of flight, $t_{T}$, and
verifies the relation $2\overline{O_{1} O_{2}}=c t_{T}$. If  the
speed of light is the same in both directions,   the beam is
reflected by  $O_{2}$ at the time $t_{R}=t_{T}/2$. This data, once
communicated from $O_{1}$ to $O_{2}$, can be  used by $O_{2}$ to
synchronize his clock with $O_{1}$'s (the Einstein procedure of
synchronization). In other words the assumption of isotropy leads
to the conclusion that the events A and R are simultaneous, so
that the definition of simultaneity suffers from the same
conventionality content of the isotropic assumption.

It is often noticed by some authors that this conclusion cannot be
drawn  essentially for two reasons.
\begin{itemize}
 \item There may be some way to synchronize distant clocks without using the isotropic assumption,
  for instance, with a slowly transport of a third clock from $O_{1}$ to $O_{2}$.
\item In the above argument we used only one experimental fact, that the speed of light
as measured over a closed path is always $c$, (hereafter this
law\footnote{Elsewhere \cite{AV} it is called "Weyl's {\em
Erfahrungstatsache}", and it is distinguished from the Reichenbach
round-trip axiom which states the independence of the round-trip
time from the direction of the journey.} is referred to, in short,
as the ``$L/c$ law''). Other experimental facts could restrict the
allowed values of the speed of light in one direction, eventually
leaving us with only the isotropic possibility.
\end{itemize}
 However, the  solution suggested in the first point simply replaces the isotropic
 convention with other equivalent assumptions \cite{ED,MS,AV}. It does not exclude the
 possibility of alternative anisotropic choices. As we shall see, the second argument does not
 work as well, because there are  anisotropic conventions  compatible with every physical law,
 the only price to be paid being  a change in their mathematical expression. Moreover, conventions
 different  from the isotropic one can prove to be natural in some contexts such as when  the
 observers live over the surface of a spinning  planet (Section \ref{pianeta}).

As a first example of an alternative convention let us return to
the one dimensional case. Following the supporters of the
conventionality of simultaneity, we are able to fix the time
reflection, $t_{R}=\epsilon t_{T}$, where $\epsilon$ is the
Reichenbach coefficient  usually taken in the range $\epsilon \in
(0,1)$. Once the choice has been made, the speed of light directed
right becomes $c/2\epsilon$ and the one directed left  becomes
$c/2(1-\epsilon)$. Whatever the choice of $\epsilon$ is, the $L/c$
law is satisfied. The restriction to the one dimensional case
however does not clarify the problem, nor exhibits the richness of
the possible conventions. Our analysis starts in the following
section where we skip to the three dimensional case.\\

\section{ANISOTROPY AND INHOMOGENEITY}
Let us consider a Euclidean space $E^{3}$ where light propagates
on straight lines. A beam of light leaves its starting point
$O_{1}$ and through the reflection over suitable mirrors covers a
closed path $\gamma$ ending in $O_{1}$. If we use a large number
of mirrors the path can approximate, as much as we want,  a smooth
closed curve of arbitrary shape, so that we can assume $\gamma$ to
be an arbitrary differentiable closed curve. If $L$ is the length
of the curve, by the  $L/c$ law, the total travelling  time is
$\tau=L/c$. Let us introduce a field $\vec{A}(\vec{x})$ so that
$\nabla \!\!\times \!\!\vec{A}=0$ (or, which is the same,
$\vec{A}=\nabla \phi$ for a suitable scalar function
$\phi(\vec{x})$), then
\begin{equation}
\tau=L/c+\oint_{\gamma}\vec{A} \cdot \dd \vec{l}.
\end{equation}
The previous expression can be rewritten
\begin{equation}
\tau=\oint_{\gamma}  \dd l  (\frac{1}{c}+A\cos \theta )=\oint_{\gamma}\frac{\dd l}{v(\vec{x}, \theta)},
\end{equation}
where $\theta$ is the angle between $\vec{A}$ and $\dd \vec{l}$ and where
\begin{equation} \label{speed}
\vec{v}=\frac{c \hat{v}}{1+c \hat{v} \cdot \nabla \phi(\vec{x})}
\end{equation}
is a new modified speed of light. It is anisotropic, in fact its
absolute value
 \begin{figure}
\centering
\includegraphics{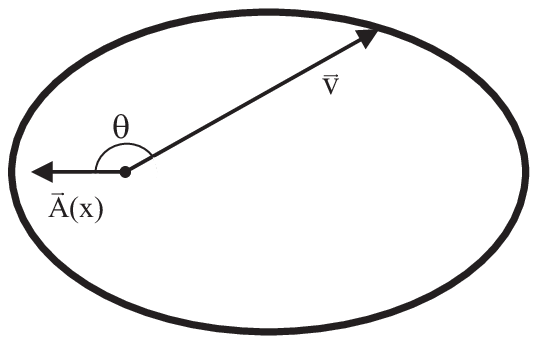}
\caption{The anisotropy of the speed of light at the point
$\vec{x}$ is elliptical.} \label{baobab}
\end{figure}
depends on the direction $\hat{v}$ (see figure \ref{baobab}). Now
it is clear that, if the speed of light has the anisotropic and
inhomogeneous value given by Eq. (\ref{speed}), then the $L/c$ law
is fulfilled. Therefore, the arbitrariness of the speed of light
amounts at least to an entire field $\phi(\vec{x})$.

There are more general expressions which verify the $L/c$ law for
any closed path starting at $O_{1}$. However, if the $L/c$ law is
verified by any observer, that is for any choice of $O_{1}$, then
the most general expression for the velocity is given by Eq.
(\ref{speed}).
 This is  proved in the
appendix, and is correctly stated by the following \\

{\bf Theorem}. Let  $M=E^{3} \times \mathbb{R}$ be a spacetime
consisting of a Euclidean space $E^{3}$ endowed with a global time
$\mathbb{R}$.  Suppose that any observer at rest measures with his
clock, and along his worldline, a time which differs from the
global time $t$ only by an additive constant. Suppose moreover
that light propagates along straight lines  with a finite speed of
norm
 $f$,
\begin{equation}\label{effe}
\vec{v}=\frac{\dd \vec{x}}{\dd t}=f(\hat{v}, \, \vec{x}, \,t)\,
\hat{v}.
\end{equation}
Finally, assume that two light worldlines coincide if they pass
through the same event with the same direction.
 Then, if the
$L/c$ law holds:
\begin{itemize}
\item[a)]  In $M$
 there is
 a  global time variable $\eta$ that  makes the speed of light isotropic
\begin{equation}
\vec{w}=\frac{\dd \vec{x}}{\dd \eta}=c \hat{w}.
\end{equation}
The function $\eta(\vec{x}, t)$ is unique up to an additive
constant, and of the form
\begin{equation} \label{trasf}
\eta=t-\phi(\vec{x})
\end{equation}
\end{itemize}
\begin{itemize}
\item[b)] If $f$ is continuous in its arguments, the function $\phi(\vec{x})$ is differentiable,
 with $|\nabla\phi(\vec{x})|<1/c$, and the
speed of light is  given by Eq.~(\ref{speed}).
\end{itemize}
Moreover, if light is, for any given direction, the fastest
signal\footnote{ Experimentally it is possible, for any given
direction, to establish unambiguously which of two signals is the
fastest. Let the two signals start at the same instant from
$O_{1}$ and reach $O_{2}$. Clearly, the first of the two signal
that reach $O_{2}$, as measured by  $O_2$'s clock, is the fastest.
This conclusion does not depend on the synchronization of the
clocks of $O_1$ and $O_2$.} that carries information then:
\begin{itemize}
\item[c)] $M$ has a Minkowskian causal structure.
\end{itemize}

 In the proof we construct, using the Einstein
 synchronization convention, a new global time variable $\eta$ which makes the speed of light
  isotropic. Then we prove that $\eta$, like $t$,  measures the time of clocks at rest (Eq.
 \ref{trasf}). The other statements follow a).

We notice that the first part of the theorem (firsts two points),
applies to whatever signal propagates over straight lines such as,
for example,  the sound. In that case, with  $c$ we mean the {\em
two-way} velocity of the signal.

  Statement a) of the theorem is in some respect opposite to
statement b). The latter implies the conventionality of the
isotropic assumption whereas the former shows that all the freedom
in choosing the convention can be eliminated with a time
coordinate change.
 This  last circumstance makes the one to one correspondence between the convention in
the velocity distribution and the convention in the concept of
simultaneity, clear. Two different time variables $\eta$ and $t$,
related by Eq. (\ref{trasf}), have indeed, different simultaneity
slices.

The theorem, moreover, suggests to replace the postulate of
constancy of the speed of light with the $L/c$ law which is
independent from the synchronization convention used.

 Anderson and Stedman \cite{AS}, starting from the isotropic convention and
making the substitution~(\ref{trasf}), obtained the conventions
expressed by (\ref{speed}). We have shown that even the converse
is true: every convention for the velocity of light, allowed by
the $L/c$ law, takes that form and, hence, every allowed
convention can be seen as arising from the isotropic convention by
a change in the clocks synchronization. A step in this direction
was already made by Anderson et al. \cite[p. 132]{AV} who proved
that such a coordinate change exists, provided the $L/c$ law
 holds,  and the anisotropy is elliptical \cite[p. 127]{AV}. However, the strong requirement of
 ellipticity was made  {\em a priori}, so that a  proof
 of our theorem was still lacking.

We have derived the causal structure of $M$ without the need of a
spacetime metric. Let, now, $M$ be a Lorentzian manifold where
light rays propagate over null geodesics. By the theorem, its
metric $\dd s^2$, in the coordinates $\{ \vec{x} , \eta \}$, must
equal Minkowski's up to a conformal factor (see, for instance,
\cite{HE}). If additionally, the metric is required to measure the
proper time of clocks at rest ($\dd s = \dd \eta$), then the
conformal factor is fixed to unity and the Minkowski metric is
completely recovered.

Finally, the reader can verify, following the proofs of the
appendix,  that the conditions of the theorem can be considerably
weakened allowing a generalization of the first part. The space
$E^{3}$ may be replaced with a Riemmanian space of
$\mathbb{R}^{3}$ topology where light is no longer required to
propagate along the geodesics of the space, but needs only to move
in trajectories invariant under inversion of direction. Then in
Eq. (\ref{speed}) with $\vec{v}$ we mean $\dd x^{i}/\dd t$ and
with $\hat{v}$ we mean $\dd x^{i}/ \dd l$ where $\dd l^{2}=h_{ij}
\dd x^{i} \dd x^{j}$ is the metric of the Riemmanian space. Here,
we are mainly concerned with the theoretical and experimental
foundations of Minkowski spacetime, so as not to discuss this
generalization further.\\

\section{CONSISTENCE OF ANISOTROPIC CONVENTIONS}
We have seen that any possible convention is related to the
isotropic one by the coordinate transformation of Eq.
(\ref{trasf}). This is the fundamental ingredient which allows us
to prove our inability to find some  physical phenomena ruling out
one convention instead of another, as shown in great detail by
Anderson  et al. \cite{AV} . Let us shortly review their findings;
this will justify the use of the word  ``convention'' which, in
the present paper, is referred to any scheme allowed by
experience.

 We can express all the known laws of physics  in a conventional global
 time coordinate obtaining a set of physical laws coherent with experience.
 The set is the one we had developed if, in our history of science, instead
 of using the isotropic convention we had chosen an anisotropic one. This set
 is recovered simply by performing a coordinate transformation from the coordinates
 of the isotropic convention, $\{  \vec{x}, \eta \}$, to the coordinates of an anisotropic
 convention,   $\{  \vec{x}, t \}$. For instance, the Gauss law of electromagnetism is written
 in the anisotropic coordinates
\begin{equation} \label{gauss}
\nabla \cdot \vec{E}+\nabla \phi \cdot \frac{\p \vec{E}}{\p t}=4 \pi \rho.
\end{equation}
As another example, the velocity of a particle of worldline $\vec{x}(t)$ is given by
\begin{equation} \label{bah}
\vec{v}=\frac{\dd \vec{x}}{\dd t}=\frac{\vec{w}}{1+\vec{w} \cdot \nabla \phi}, \qquad \textrm{ where} \ \vec{w}=\frac{\dd \vec{x}}{\dd \eta},
\end{equation}
and the proper time of the particle is
\begin{equation} \label{slow}
\dd \tau=\sqrt{1-\frac{\vec{w}^{2}}{c^{2}}} \,\dd \eta=\sqrt{(1-\vec{v} \cdot \nabla \phi)^{2}-\frac{\vec{v}^{2}}{c^{2}}} \, \dd t .
\end{equation}
Not every physical law requires a time variable for its
formulation. When it is possible, a convention-free formulation
 clearly reveals the physical meaning of the law. We have already shown
 that the causal structure is independent
from the chosen convention. In $M=E^{3} \times \mathbb{R}$ we can
even define, in a convention-free manner, the ``light cone'' of an
event A,
\begin{center}
{\em An  event B is said inside the light cone of an event A if
there is  a path $\gamma$} \\ {\em  which allows a light beam
starting in A, and travelling over $\gamma$, to end in B.}
\end{center}
As a consequence, the law which states that  information cannot
propagate faster than light, can be expressed in a convention-free
way, too.

In the literature there are theoretical   proofs \cite{JA} of the
isotropy of the speed of light but all of them use a non-modified
law of physics in the process, and no one deals with the problem
of velocity inhomogeneity. It is an easy task to ``prove'' the
isotropy of the speed of light if we implicitly use an assumption
or a law that holds only in the isotropic convention. For
instance, it is easy to prove the isotropy of the speed of light
if we use the Gauss law in its standard form. Indeed, its
predictions agree with physical phenomena, Eq.~(\ref{gauss}), only
if the velocity of light is isotropic. In the same way, contrary
to the claims of Will \cite{WI}, it is impossible to
experimentally test the isotropy of the speed of light. Many of
the experiments he had analyzed have some value as tests of
special relativity, but as tests of the isotropy of the speed of
light, they are illusions \cite[p. 148]{AV}.\\

\section{SIMULTANEITY FROM CAUSALITY}
In the previous section, we have shown that  the laws of physics
are simplified in the Einstein convention whereas, in anisotropic
conventions, they loose their symmetries. The requirement of some
symmetry becomes a way to restrict the allowed conventions to the
isotropic one. This is seen even in the causal approach to the
conventionality of simultaneity \cite{GR}  whose cornerstone is
the theorem of Malament \cite{MA}.  This theorem (see \cite{Va}
for a readable account of it)  essentially proves that, if the
causal structure of spacetime is that of Minkowski, and $C$ is the
wordline of an observer at rest, the only non trivial equivalence
relation (simultaneity relation) invariant under $C$-causal
automorphisms (diffeomorphisms of spacetime that preserve causal
relations and  map  $C$ onto itself) is that of Einstein. Our
theorem enforces that of Malament in the following sense. The
causal structure of spacetime is non-conventional because it can
be tested experimentally. However, Malament takes it for granted
that it is Minkowskian, that is, derived from  a pseudo-Riemmanian
manifold of $\mathbb{R}^{4}$ topology, where the metric
\begin{equation} \label{mink}
\dd s^{2}=c^{2} \dd t^{2}-\dd \vec{x}^{2}
\end{equation}
vanishes on light worldlines. Now, one may wonder if this belief
is well founded. After all, from (\ref{mink}), there also follows
 that the speed of light is isotropic (by dividing by $\dd
t^{2}$). In other words we cannot rely on equation (\ref{mink}) to
state the causal structure of spacetime because it is only
compatible  with the isotropic assumption. We need experimental
evidence that the causal structure of spacetime is  Minkowskian
and to do this we cannot rely on speed of light conventions.
Alternatively, we need a proof of the independence of the causal
structure from the  convention chosen.  At this point, the last
part of our theorem enters.
 It states that, because of the $L/c$ law,  the
causal structure of spacetime is  Minkowskian even if our
spacetime $M=E^{3} \times \mathbb{R} \,$ is not a Lorentzian
manifold\footnote{Remember that the causal structure of spacetime
follows  our theorem without the need of a spacetime metric.
Moreover,  Malament's argument uses only the causal structure of
Minkowski spacetime.}. Malament's argument then works. We
summarize the entire deduction in  a scheme
\begin{equation*}
\boxed{M=E^{3} \times \mathbb{R} \ \xrightarrow{L/c \ \text{law}}  \substack{\text{{\normalsize Minkowskian}} \\ \text{{\normalsize causal structure}}}   \xrightarrow{\substack{\text{Malament's} \\ \text{argument}} } \textrm{Einstein convention.} }
\end{equation*}
Although  attractive, in what  follows we  abandon this approach
to the  conventionality of simultaneity essentially for one
reason. Malament, in order to recover the Einstein convention,
requires an invariance principle, namely the invariance of the
simultaneity relation under C-automorphisms; but we have seen that
a number of physical laws have the same effect if we require some
symmetry. There is no physical reason for such a requirement;
after all, the concept of simultaneity has to do with clocks not
with light (causal structure) and in this regard clocks say that
there are a number of viable conventions, those given by Eq.
(\ref{speed}). Moreover, Malament's argument is hard to generalize
to the case of observers in generic motion \cite{BU}, or to
generic spacetimes, because in such contexts C-automorphisms may
be absent.\\

\section{THE CHOICE OF A GOOD CONVENTION}
Once one agrees on the conventional nature of  simultaneity, the
problem becomes how to find a good convention for the physical
context at hand. We suggest three criteria
\begin{itemize}
\item Simplicity of the laws of physics.
\item Invariance of the convention used under  change of the observer.
\item Existence of a global time variable.
\end{itemize}
The first criterion is clear, one has to choose the simplest
convention whenever the last two points are satisfied. The second
criterion has the following meaning: a convention is good if it is
the same for every observer, in such a way that  communication
among them is possible without referring each time to a subjective
choice. This implies that the function $\phi$ must be the same for
all observers: indeed, if the observer $O_{1}$ uses the time
variable $t_{1}=\eta_{1} -\phi_{1}$ and the observer $O_{2}$ uses
the time variable $t_{2}=\eta_{2} -\phi_{2}$, a communication
among them is useless, unless each observer knows the function
$\phi$ used by the other. To meet the second point, the set of
observers must agree that function $\phi$ be used.  Depending on
the physical situation, we have to restrict and define the set of
observers under which the invariance of the convention holds. This
happens in the following example.

So far we have considered only observers at rest, here we look for
conventions well suited for moving observers.
 In this example, our set of observers, under which invariance of the convention must be met, is given by inertial observers.
Let us consider the Galilean principle of relativity
\begin{center}
  (*) $\qquad$ {\em A   reference  frame  in  uniform  rectilinear  motion  with  respect \\
to  an  inertial  frame  is  inertial  by  itself.}
\end{center}
Here, for ``inertial frame'', we mean any observer who does not
feel inertial forces. This definition, based on detectable forces,
avoids any convention and is ideal for our purpose. The function
$\phi$, common to all inertial observers, is taken in such a way
that the  Galilean relativity principle  (*) remains unchanged
passing from the time variable $\eta$ to the time variable $t$.
This restricts the allowed conventions to a subset where the
relativity principle holds and where the laws of physics happen to
be particularly simple.  Let $\vec{w}$ be the uniform  velocity of
an inertial observer,  from Eq. (\ref{bah}), we see that in the
time variable $t$ the inertial observer has a uniform velocity
only if $ \nabla_{\vec{w}}  \nabla_{\vec{w}} \phi=0 $  which
implies\footnote{Torretti \cite{TO,T2}, already considered this
convention for a single observer requiring that Newton's first law
be satisfied.}   $\phi=\vec{a} \cdot \vec{x} +\textrm{const.}$.
With this choice, the coordinate transformations from one inertial
observer to another form a group. If   $U(a)$ is the coordinate
transformation from $\{ \vec{x}, t \}$ to $\{ \vec{x}, \eta \}$
then the coordinate transformation to a second observer of
velocity $\vec{v}$ is given by \cite{AV,UN}
\begin{equation}
G(\vec{a}, \vec{v})= U(a)^{-1} \Lambda(\vec{w}) U(a)
\end{equation}
 where $\vec{v}$ and $\vec{w}$ are related by Eq. (\ref{bah}).
 The group of coordinate transformations can be shortly written $G= U(a)^{-1} \Lambda \, U(a)$
 where $\Lambda$ is the Lorentz group in the usual representation.
We take, $\vec{a}=\frac{\alpha}{c} \hat{i}$, where $\alpha$ is a dimensionless constant.
In  one spatial dimension  the modified Lorentz transformation becomes
\begin{eqnarray}
x' & = & \gamma(w(v)) \{ (1+\alpha \beta)x-\beta c t \}        \\
c t' & = & \gamma(w(v)) \{ (1-\alpha) \beta ct +(1+\alpha-\alpha \beta+\alpha^{2} \beta )x \},
\end{eqnarray}
where
\begin{eqnarray}
\beta & = & \frac{w}{c}=\frac{v/c}{1-\alpha v/c}, \\
\gamma(w(v)) & = & \frac{1-\alpha v/c}{\sqrt{(1-\alpha v/c)^{2}-(v/c)^{2}}}.
\end{eqnarray}
With the simplest choice,  $\alpha=0$, we recover the Lorentz transformation.

We mention another interesting convention. Starting from a
realistic viewpoint, and with purposes very different from ours
\cite{S2}-\cite{R2}, Selleri \cite{SE} renewed some interest in
the absolute synchronization  convention \cite{MS}, that is on the
proposal $\phi_{\vec{v}}= (\vec{v} \cdot \vec{x})/c^{2}$, where
$\vec{v}$ is the velocity of the observer $O_{\vec{v}}$ with
respect to a privileged frame $O_{0}$. This convention, depending
on the velocity $\vec{v}$ of the observer, is not invariant under
change of inertial frame. As a consequence, the laws of physics
are not invariant  either and the modified Lorentz transformations
do not form a group. However, the Galilean relativity principle
(*) remains true because $\phi_{\vec{v}}$ is linear in $\vec{x}$.
Implicitly, in the previous section, much in the spirit of
Mansouri and Sexl \cite{MS},   a problem raised by Selleri in his
paper \cite{SE} is solved, that of finding how the laws of physics
must be written in the absolute synchronization convention and if
there is an experiment capable of ruling it out \cite{CS}.
Nevertheless, we stress, in contrast to him \cite{S2}, that the
possibility of anisotropic conventions does not imply the
inconsistency of special relativity.

The third point requires a wider discussion; we devote the
following section to it.\\

\section{THE EXISTENCE OF A GLOBAL TIME VARIABLE} \label{pianeta}
So far, we have considered only the case in which the $L/c$ law
holds everywhere; to generalize our treatment the field
$\vec{A}(\vec{x})$ is now taken arbitrarily, that is, we remove
the condition $\vec{A}=\nabla \phi(\vec{x})$. If light has the
velocity
\begin{equation} \label{new}
\vec{v}=\frac{c \hat{v}}{1+c \hat{v} \cdot \vec{A}(\vec{x})},
\end{equation}
then  the time taken by a light beam to travel  round trip over
the path $\gamma$ is given  by
\begin{equation}
\tau=L/c+\oint_{\gamma}\vec{A} \cdot \dd \vec{l},
\end{equation}
and the difference from the case in which light travels in the
opposite sense is
\begin{equation} \label{sagnac}
\delta \tau=2 \oint_{\gamma}\vec{A} \cdot \dd \vec{l}.
\end{equation}
This is a generalized Sagnac effect due to the distribution of
velocities,~Eq.~(\ref{new}). Being a measurable quantity, if a
Sagnac effect is present, every allowed convention must account
for it. Hence, in the presence of a Sagnac effect, it is
impossible to find a new global time variable which allows the
speed of light to be everywhere $c$. The existence of a global
time variable is very useful, and must be considered  one of the
main tasks of a good convention, so this excludes the isotropic
convention in a number of physical situations. Moreover, if two
conventions on the velocity of light are allowed, because they
lead to the same correct predictions (Sagnac effect), then their
fields are linked to one another by the gauge transformation
\begin{equation} \label{gauge}
\vec{A}(\vec{x}) \, \longrightarrow \, \vec{A}(\vec{x})+ \nabla \phi(\vec{x}).
\end{equation}
In order to see this, take the difference of two $\vec{A}$ fields,
then obtain a new  field that is rotation-free because its
integral over an arbitrary  closed path is zero. Measurable
quantities, being convention-free quantities, must be gauge
invariant. For instance, the rotation $\vec{B}(\vec{x})= \nabla
\times \vec{A}$ is gauge  independent and plays the role of the
field strength of the gauge theory. It can be measured revealing
the Sagnac effect for a closed path in the neighborhood of
$\vec{x}$. Again, the gauge transformation (\ref{gauge}) follows
from a resynchronization of clocks, that is from a  transformation
in the global time coordinate $t \rightarrow t+\phi(\vec{x})$.

Our treatment generalizes  Anderson and Stedman's \cite{AS} who
presented the choice of simultaneity, within an inertial frame, as
a gauge choice.

In our formalism, the usual non-relativistic Sagnac effect
\cite{SA,ST} is obtained once one takes
\begin{equation} \label{sagnac2}
\vec{A}(\vec{x})=\frac{\vec{\omega} \times \vec{x}}{c^{2}},
\end{equation}
where $\omega$ is the angular velocity of the rotating platform.
Indeed, with this choice, one recovers the well known formula
\begin{equation} \label{sagnac3}
\delta \tau= \frac{4 S \omega}{c^{2}},
\end{equation}
where $S$ is the area of the surface subtended by the curve
$\gamma$. Eq. (\ref{sagnac2}) is even the best convention that
people living on the Earth surface can take, where $\vec{x}$ is
the displacement from the Earth's axis.

Eq. (\ref{sagnac2}) can be recovered from the  general
relativistic work of M{\o}ller \cite{MO} who proved that an
expression like (\ref{new}) with\footnote{Indices are lowered and
raised with the spatial metric  $
h_{ij}=g_{ij}-g_{0i}g_{0j}/g_{00}$.} $A_{i}=g_{0i}$ holds  in
whatever  stationary frame  in which $g_{00} = -1$,  where
$x^i=const.$ is the worldline of a generic observer at rest in the
frame. In our case \cite{MO},
$g_{00}=-(1-\frac{r^2\omega^2}{c^2})$, and the previous condition
holds, for a given radius, only in the non-relativistic limit. In
more general relativistic circumstances, $g_{00}\ne -1$, the
chosen global time variable no longer measures the rate of clocks
at rest, and Eq.~(\ref{new}) no longer holds. However,  using
fiber bundle techniques one can still reveal  the gauge nature of
simultaneity, as we shall show in a forthcoming paper.

In the present paper we are mainly concerned with the case of a
Euclidean space in which the  $L/c$ law holds, at least locally.
In our theorem we proved that  $\vec{A}=\nabla \phi$. However, if
the  $L/c$ law does not hold globally, we can only conclude that
$\nabla \times \vec{A}=0$. In a space not simply connected  this
does not leave out the possibility of a Sagnac effect, and, as a
consequence, this proves that the isotropic convention can be
unsuitable even when the $L/c$ law locally holds.

Think, as an example, of a large cylindrical spaceship of radius
$R$ spinning along its axis at  angular velocity $\omega$. People
live on the internal cylindrical surface of the spaceship and
$\omega$ is chosen to reproduce the gravitational acceleration $
g=\omega^{2} R$. Light and electric signals propagate along the
surface. In such a situation  the  $L/c$ law holds locally but a
Sagnac effect is present when a light beam travels all around the
spaceship.\\

\section{CONCLUSIONS}

In the first part of the paper we  developed the consequences of
the $L/c$ law. We found that the allowed  distributions for the
velocity of light are given by Eq. (\ref{speed}), and that each of
them can be recovered from the isotropic value via a time
coordinate transformation. The relation with a coordinate
transformation enabled us to rewrite the laws of physics
coherently with the anisotropic convention adopted. This change in
their expression clearly does not alter the physical content of
the laws, so that a physics based on anisotropic conventions
appears to be feasible. Moreover,  anisotropic conventions can
prove extremely useful, as we  showed in the last part of the
paper. The approach to the conventionality of simultaneity  as a
gauge theory seems very attractive and will be the subject of
subsequent works. It can be considered  a step towards general
relativity.

Our analysis of the $L/c$ was proved useful even in the causal
approach to the conventionality of simultaneity. We showed that,
if the $L/c$ law holds in  a Euclidean space, then the associated
spacetime is causally the same as Minkowski spacetime. This result
relates convention-free concepts, namely the $L/c$ law and the
causal structure of spacetime. It justifies Malament's argument if
one wants to base the simultaneity concept on the causal
structure. This last approach, however, appears untenable when one
skips from Minkowski spacetime to more realistic spacetimes where
causal automorphisms are absent.\\

\section*{ACKNOWLEDGMENT}
I wish to thank Bruno Carazza for his kind suggestions and for
having shared useful discussions.\\

\section*{APPENDIX}
The theorem requires a lemma. \\
{\bf Lemma.} Under the hypothesis of the theorem, let $O_{1}$ be
an observer at rest.  Let  $t_{1}$ be the time measured by the
clock of $O_{1}$ along its own worldline $\vec{x}_{O_1}=const.$
From the hypothesis, on the worldline of the observer, we have
$t_{1}=t+const.$ where $t$ is the global time. Outside the
wordline of the observer $O_1$ the parameter $t_1$ is not yet
defined. If  $O_{1}$ labels events outside his worldline using the
Einstein synchronization convention, then the variable $t_{1}$
becomes a global function: $t_{1}=\phi_{1}(\vec{x}, t)$. Moreover,
$\phi_{1}$ turns out to be increasing and continuous in $t$. For a
given $\vec{x}$, $t_1$ takes any real value and the speed of
light, in this new variable, becomes $c$: $\dd \vec{x}/ \dd t_{1}=
c \hat{v}$.
\begin{figure}[!ht]
\centering
\includegraphics{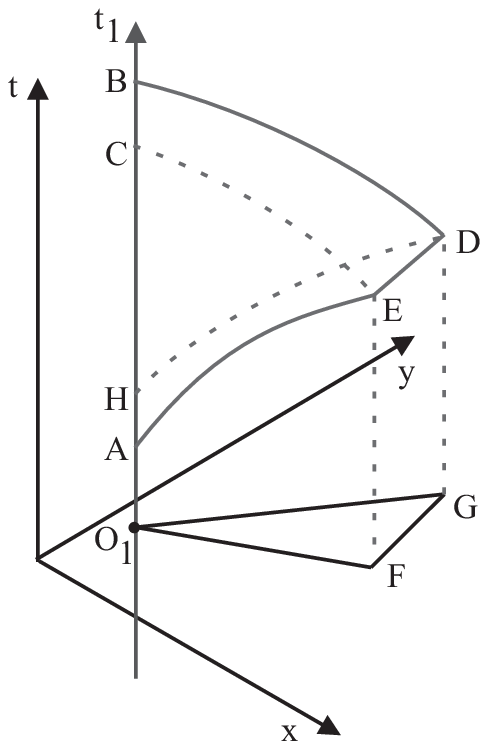}
\caption{Note: the worldline of the light beam is not necessarily straight.}
\label{figura4}
\end{figure} \\
{ \bf Proof of the lemma}\footnote{This lemma can be generalized
in the case of   curved spaces. You have simply to read
$\overline{AB}$ as the length of the trajectory of the light beam
from $A$ to $B$.}.  Look at figure \ref{figura4} where the
coordinate $z$ is omitted for the sake of clarity. Starting from
$O_{1}$, and being reflected in $F$ and $G$, a light beam travels
all along the path $O_{1} F G O_{1}$. $F G$ is an infinitesimal
displacement and our task is to show that the speed of light over
$FG$  is $c$ if the time used is $t_{1}$. We added to the picture
the trajectories of the light rays $CE$ and  $HD$ useful to define
$t_{1}(D)$ and $t_{1}(E)$. From our hypothesis
\begin{eqnarray}
c(t_{B}-t_{A}) & = & \overline{FO_{1}}+\overline{FG}+\overline{O_{1}G} \\
c(t_{C}-t_{A}) & = & 2 \overline{FO_{1}} \\
c(t_{B}-t_{H}) & = & 2 \overline{O_{1}G} \\
\Rightarrow \  \overline {FG} & = & \frac{c}{2}(t_{B}+t_{H}-t_{A}-t_{C})=c(t_{1}(D)-t_{1}(E)).
\end{eqnarray}
Or, more explicitly
\begin{equation}
\frac{\dd t_{1}}{\dd l}=\frac{1}{c} \, .
\end{equation}
For a given $\vec{x}$, $t_1=\phi_1(\vec{x}, t)$ takes any real
value. To see this, integrate the previous equation over closed
paths of arbitrary length starting from $\vec{x}$. \\
The  definition of $\phi_1$ implies that $t_1$  increases with $t$
(because, by hypothesis, two worldlines with the same direction
cannot intersect); hence $\phi_1(\vec{x}, t)$ is invertible and
$t_1$ can be used to label events. Finally $\phi_1$ is continuous
in $t$ because, for a given $\vec{x}$, it is increasing in $t$ and
its image is $\mathbb{R}$.
 The lemma is proved. \\
 Notice that we have not yet shown that the "time" function
$t_{1}=\phi_{1}(\vec{x}, t)$ measures  the flow of time for observers different from $O_{1}$. We are ready to prove the theorem. \\
{\bf Proof of the theorem.} Let us prove that a global time
variable $\eta$ which makes the speed of light isotropic must be
unique up to an additive constant. Let $\gamma$ be the worldline
of a light beam of direction  $\hat{v}$. If $\eta_1$ makes the
speed of light isotropic
\begin{equation}
\frac{\dd \eta_1}{\dd l}=\frac{1}{c},
\end{equation}
where $l$ is the natural parameter of the projection of $\gamma$
on $E^3$. Subtracting this equation with the same equation for
$\eta_2$ we find that $\eta_1=\eta_2+c(\gamma)$ for all the events
that lie on the worldline $\gamma$ of the light beam considered.
The constants of two different  light beams must be equal if their
worldlines intersect. In the coordinates $\{\vec{x}, \eta_1\}$ the
light cone
 of an event $a=(\vec{x}(a), \eta_1(a))$ has the usual equation
$ |\vec{x}-\vec{x}(a)|=|\eta_1-\eta_1(a)| $, therefore the light
cones of two events $a$ and $b$, intersects. This implies that a
light worldline passing through $a$ intersects  a light worldline
passing through $b$ and hence that
\begin{equation}
\eta_1(a)-\eta_2(a)=\eta_1(b)-\eta_2(b)
\end{equation}
Making $b$  arbitrary, we see that a constant $c$ exists such that
$\eta_1=\eta_2+c$.

Let $t_{1}=\phi_{1}(\vec{x}, t)$ and $t_{2}=\phi_{2}(\vec{x}, t)$
be the functions of the lemma related to the observers $O_{1}$ and
 $O_{2}$. These functions are global time variables which make
 the speed of light isotropic. From our previous result $t_1$
 and $t_2$ differ by an additive constant. But $t_2$, over the
 worldline $\vec{x}=\vec{x}_{O_2}$, equals the  global time
 variable $t$ up to an additive constant and  hence, because $O_2$
 is arbitrary,
\begin{equation} \label{due}
\phi_{1}(\vec{x}, t)=t-\tilde{\phi}_{1}(\vec{x})
\end{equation}
for a suitable scalar function $\tilde{\phi}_{1}(\vec{x})$. This
proves a).

Let now $f$ be continuous in its arguments. By hypothesis the
worldline $\vec{x}(t)$ of a light beam of direction $\hat{v}$ is
derivable (see Eq. (\ref{effe})).  This implies that $t(l)$ is
derivable in  the open set $A \subset S^2 \times M$ where
$f(\hat{v}, \vec{x}, t)
> 0$. The lemma shows  that $t_1(l)$ is derivable too, therefore  the same is true for $\tilde{\phi_1}(\vec{x}(l))$ and we obtain
\begin{equation} \label{derivata}
\nabla_{\hat{v}}\tilde{\phi}_{1}(\vec{x})=\frac{1}{f(\hat{v},\,
\vec{x},\, t)}-\frac{1}{c} \quad \text{in} \ A.
\end{equation}
This proves that $f$ is independent of time in $A$. Considering
only the $t$ dependence, $f$ takes  two values: zero outside $A$
and $ {c }/[{1+c \nabla_{\hat{v}}\tilde{\phi}_{1}(\vec{x})} ]$
inside $A$. However, $f$ is continuous in $t$ and hence
independent
of time throughout $S^2 \times M$.\\
If $f(\hat{v}, \vec{x})>0$ there is a neighborhood  of $\hat{v}
\in  S^2$ where we can find three linear independent vectors
$\hat{v}_1, \hat{v}_2, \hat{v}_3$ which verify $f(\hat{v}_i,
\vec{x})>0$. The partial derivatives
$\nabla_{\hat{v}_i}\tilde{\phi}_{1}(\vec{x})$ are continuous in
$\vec{x}$ (see Eq. (\ref{derivata})) and therefore
$\tilde{\phi}_{1}$ is differentiable: $
\nabla_{\hat{v}}\tilde{\phi}_{1}(\vec{x})=\hat{v} \cdot
\nabla\tilde{\phi}_{1}(\vec{x})$.\\
Let $B \subset S^2 \times E^3$ be the open subset where
$f(\hat{v}, \vec{x})>0$. From Eq. (\ref{derivata}),
\begin{equation} \label{s2}
f(\hat{v}, \vec{x})=\frac{c }{1+c \hat{v} \cdot
\nabla\tilde{\phi}_{1}(\vec{x})} \quad \text{in} \ B,
\end{equation}
therefore $f \ge {c }/[{1+c | \nabla\tilde{\phi}_{1}(\vec{x})|}]$
in $B$. If $(\hat{v}_1, \vec{x}_1) \in B$ then $\forall$  $\hat{v}
\ (\hat{v}, \vec{x}_1) \in B$, indeed, for a given $\vec{x}_1$,
there is no continuous function $f: S^2 \rightarrow \mathbb{R}$
which is greater than $ {c }/[{1+c |
\nabla\tilde{\phi}_{1}(\vec{x}_1)|}]$ in an open
subset of $S^2$ and zero elsewhere.\\
Let $C \subset  E^3$ be the open subset where $f(\hat{v},
\vec{x})>0$. From Eq. (\ref{s2}), because $f$ is finite,
$|\nabla\tilde{\phi}_{1}(\vec{x})|<1/c$ or $f(\hat{v},
\vec{x})>c/2$ in $C$. But $f$ is continuous in $\vec{x}$
throughout $E^3$ therefore $C=E^3$, $B=S^2 \times E^3$ and $A=S^2
\times M$. In other words $f$ is positive.  Eq. (\ref{s2}) proves
statement b).

In order to prove c) notice that the causal structure does not
depend on the coordinates used to label events. From a) and using
the coordinates $\{\vec{x}, t_1\}$ we recover the light cone
structure of Minkowski spacetime, and since, by hypothesis, light
is the fastest signal, the causal structure of $M$ coincides with
that of Minkowski spacetime.\\

\end{document}